\documentclass[preprint]{elsarticle}

\usepackage{subfigure}
\usepackage{epstopdf}

\usepackage{multirow}
\usepackage{amsmath}
\usepackage{graphicx,url}
\usepackage{verbatim}
\usepackage{amssymb}

\journal{Computer Networks}

\begin{document}

\begin{frontmatter}



\title{Server Placement with Shared Backups for Disaster-Resilient Clouds}

\author[ufrj,uerj]{Rodrigo S. Couto\corref{cor1}}\ead{rodrigo.couto@uerj.br}
\author[upmc]{Stefano Secci}\ead{stefano.secci@upmc.fr}
\author[ufrj]{Miguel Elias M. Campista}\ead{miguel@gta.ufrj.br}
\author[ufrj]{\\ Lu\'is Henrique M. K. Costa}\ead{luish@gta.ufrj.br}
\address[ufrj]{Universidade Federal do Rio de Janeiro - COPPE/PEE/GTA - POLI/DEL\\
	       P.O. Box 68504 - CEP 21941-972, Rio de Janeiro, RJ, Brazil\\
}
\address[uerj]{Universidade do Estado do Rio de Janeiro - FEN/DETEL/PEL\\
	       CEP 20550-013, Rio de Janeiro, RJ, Brazil\\
}
\address[upmc]{Sorbonne Universit\'es, UPMC Univ Paris 06, UMR 7606, LIP6 \\ 
		F-75005, Paris, France
}
\cortext[cor1]{Corresponding author.}

\begin{abstract}
A key strategy to build disaster-resilient clouds is to employ backups of virtual machines in a geo-distributed infrastructure. Today, the continuous and acknowledged replication of virtual machines in different servers is a service provided by different hypervisors. This strategy guarantees that the virtual machines will have no loss of disk and memory content if a disaster occurs, at a cost of strict bandwidth and latency requirements. Considering this kind of service, in this work, we propose an optimization problem to place servers in a wide area network. The goal is to guarantee that backup machines do not fail at the same time as their primary counterparts. In addition, by using virtualization, we also aim to reduce the amount of backup servers required. The optimal results, achieved in real topologies, reduce the number of backup servers by at least 40\%. Moreover, this work highlights several characteristics of the backup service according to the employed network, such as the fulfillment of latency requirements.\footnote{The final publication is available at Elsevier via\\ http://dx.doi.org/10.1016/j.comnet.2015.09.039} 
\end{abstract}

\begin{keyword}
cloud networking \sep resilience \sep geo-distributed data centers \sep infrastructure as a service.

\end{keyword}

\end{frontmatter}

\section{Introduction}

Many corporations are migrating their IT infrastructure to the cloud by using IaaS (Infrastructure as a Service) services.
Using this type of service, a corporation has access to virtual machines (VMs) hosted on a Data Center (DC) infrastructure maintained by the cloud provider. The use of IaaS services helps cloud clients reduce the effort to maintain an IT infrastructure; with IaaS, clients relinquish the control of their physical infrastructures. Therefore, they only rely on IaaS services if providers can guarantee performance and security levels. To encourage IaaS subscriptions, cloud providers usually try to offer high resilience levels of their VMs. To this end, IaaS providers deploy redundancy on their infrastructure to overcome various types of failures, such as hardware (e.g., failure in hard disks, network cables, and cooling systems), software (e.g., programming errors), and technical staff (e.g., execution of wrong maintenance procedures). This strategy, however, does not guarantee service availability under force majeure and disaster events that are out of the provider's control. 

Force majeure and disaster events, such as terrorist attacks and natural disasters, are situations outside of the provider's control, which can affect several network links as well as whole buildings hosting data centers. Cloud providers thus generally do not cover this type of event in their SLAs (Service Level Agreements)~\cite{couto2014Network}. Although IaaS providers often do not consider catastrophic events, they can offer recovery services such as VM replication and redundant network components to improve the resilience to clients running critical services. These services can be provided as long as a DC infrastructure resilient to disasters is available, which is generally composed of several sites spread over a region and interconnected through a wide area network (WAN)~\cite{secci2014Cloud}. Each site has a set of servers interconnected using a local network~\cite{couto2012Reliability,ferraz2014Two}.
A resilient IaaS cloud must thus employ a geo-distributed DC to eliminate single points of failure and must employ mechanisms to perform VM backups. Obviously, clients willing to have higher resilience guarantees will pay the cost of maintaining such infrastructure.

In this work, we focus on the design of disaster-resilient DCs with zero VM state loss (e.g., loss of disk and memory content) after a disaster. This means that the provider guarantees zero RPO (Recovery Point Objective) on its VMs. RPO is the time elapsed between the last backup synchronization and the instant when the disaster happens. Hence, it gives an idea of data loss after a disaster~\cite{couto2014Network}. Some critical services demand a low RPO or even zero RPO, such as banking transactions, requiring continuous data replication. 
Basically, an IaaS with zero RPO consists on VMs that continuously send backups to a server. In this case, an operation demanded by an end user is only accomplished after the VM receives an acknowledgment from the backup site, indicating that the VM state was correctly replicated~\cite{rajagopalan2012secondsite}.
As this type of service requires continuous data replication, it requires a high network capacity. Furthermore, as it needs backup acknowledgment, the primary server, i.e., the server hosting the operational VMs, and the backup one must have low latency links between each other.

The literature about resilient physical server placement considers a traditional DC distribution, such as those employed by content delivery networks (CDNs)~\cite{habib2012design,xiaoJoint2014}. In these works, the DC services are replicated through a geo-distributed architecture using anycast.
Hence, any node that runs the required services are operational and can reply the requests from clients. Consequently, the primary servers and their backups are both running at the same time.
Nevertheless, these works do not consider the synchronization of service replicas, disregarding RPO requirements. 

This work analyzes the behavior of IaaS services with zero RPO in real WAN topologies. We propose a physical server placement scheme, which designs the DC by choosing where to install the primary servers and their corresponding backups. The placement scheme has to take into account the failure model, in such a way that a disaster does not damage the primary server and its backup at the same time. In addition, the proposed scheme takes advantage of virtualization to reduce the number of backup servers. The basic idea is that a backup server needs to instantiate VMs only after a given disaster occurs~\cite{wood2010disaster}. We thus argue that, in a virtualized environment, it is inefficient to provide a dedicated backup server for each primary one. Instead, the proposed scheme aims at sharing backup servers, allowing them to receive replications from different primary servers. To share these resources, the primary and backup servers must not fail at the same time. We apply the proposed scheme in WAN topologies and show that 
backup sharing can reduce by at least 40\% the number of required servers,
as compared to the case with dedicated backups. We also quantify the capacity of each WAN topology in terms of number of primary servers supported, which directly affects the number of supported VMs. Using these results, we show that more stringent resilience requirements reduce by at least 50\% the number of primary servers supported. Our work differs from the literature by considering the service replication, which incurs in stringent latency and bandwidth requirements. In addition, the current proposals based on anycast do not save backup resources, since all backup servers are also operational. We thus focus on IaaS models, different from traditional CDNs.

This work is organized as follows. Section~\ref{zeroRPOsec:model} describes the service model and our design decisions.
Based on these decisions, Section~\ref{zeroRPOsec:problem} introduces the proposed optimization problem. Section~\ref{zeroRPOsec:eval} shows the results of the optimization problem when applied to real WAN networks. Finally, Section~\ref{zeroRPO:secRelacionados} presents related work and Section~\ref{zeroRPO:conc} concludes this work and points out future directions.

\section{Modeling and Design Decisions}
\label{zeroRPOsec:model}

The optimization problem proposed in this work distributes primary and backup servers in a given WAN topology. 
The primary servers are employed to host operational VMs, which are accessed by Cloud users through gateways spread across the WAN;
Backup servers receive VM copies from these servers. A VM backup is a complete copy of its primary VM, but it keeps in standby mode in a normal situation. 
Each primary server replicates VM copies to a single backup server installed in another DC site. This section details the DC design decisions considered in the optimization problem formulation, which is described later in Section~\ref{zeroRPOsec:problem}.

\subsection{VM Replication}

The VM backup scheme considered in this work is based on continuous and acknowledged VM replication, which allows the provider to guarantee zero RPO (Recovery Point Objective) when a disaster occurs. This type of scheme is common in local networks, being natively available in virtualization platforms such as Xen~\cite{cully2008remus}. 
More recently, VM backup schemes with zero RPO using wide area networks (WANs) started to be addressed in the literature~\cite{rajagopalan2012secondsite,wood2011Pipecloud}.
As an example we can cite SecondSite~\cite{rajagopalan2012secondsite}, employed as a reference throughout this article.
To achieve zero RPO, SecondSite is based on checkpoints. A checkpoint is defined as the VM state (e.g., disk, memory, CPU registers) at a given instant. Such state is continuously sent to a backup server that, in its turn, sends an acknowledgment to the primary server for each received checkpoint. 
The basic of operation of a VM is to run applications that receive requests from users on the Internet and reply these requests.
Before a checkpoint acknowledgment, network packets sent from the VM applications to the users are held in a queue, waiting for the upcoming acknowledgment. When the backup server confirms the checkpoint replication, all packets in the queue are sent to users.
Hence, the final user only receives a reply to his requests after the correct replication in the backup server.
Note that SecondSite imposes strict bandwidth and latency requirements. The high bandwidth utilization is due to the continuous data replication, which increases 
with the frequency of changes in the VM state.
The strict latency requirements are imposed due to the checkpoint acknowledgment before replying to users.
Hence, the lower the latency between primary and backup servers, the higher the throughput of VM applications.

When SecondSite detects a failure in the primary server, it activates the VMs stored in the backup server.
The failure is detected for each node upon the lack of replication data between servers. That is, a backup server infers that there was a failure in the primary server when it ceases to receive VM replication for a given time. In its turn,
the primary server infers that the backup is offline when it does not receive checkpoint acknowledgments for a given time. Note that both failure cases can happen not only when a given server fails, but also when the replication link is down.
Hence, if the link fails, the backup server may infer that the primary one is down and activate the backup VMs.
This can cause a problem known as split brain, where both primary and backup servers are replying to requests from clients, causing data inconsistency.
To overcome this problem, SecondSite employs another server type, called \lq\lq quorum server\rq\rq \,.
When a failure is inferred, the primary or the backup server communicates with a quorum server.
As this server exchange messages with both backup and primary servers, it can report the failure type to these servers and thus both can perform the appropriate operations. For example, if the replication link fails, the backup server is turned off. Obviously, the quorum servers should be placed on the network in such a way to quickly and efficiently detect failure.
We assume in this work that quorum servers are correctly placed and can detect all possible failures.
This type of server placement is still not addressed in the literature, but shares common aspects with SDN (Software-Defined Networking) controller placement~\cite{muller2014survivor}.

\subsection{Server Placement}

\begin{figure}
\centering
{\includegraphics[width=0.79\textwidth]{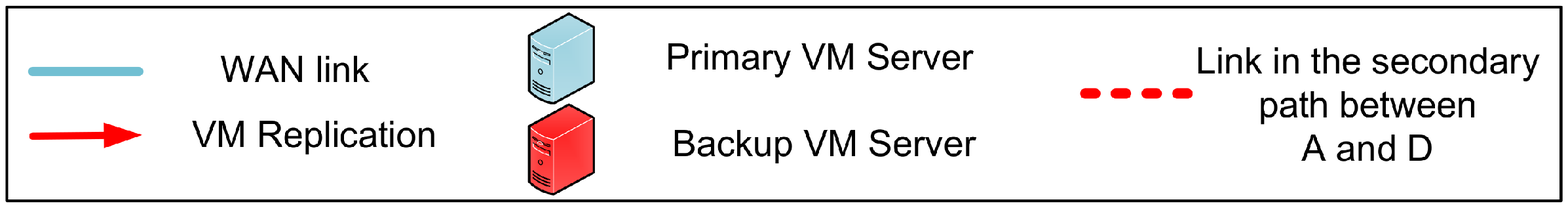}}\\
\subfigure[Server Placement.]
{\includegraphics[width=0.45\textwidth]{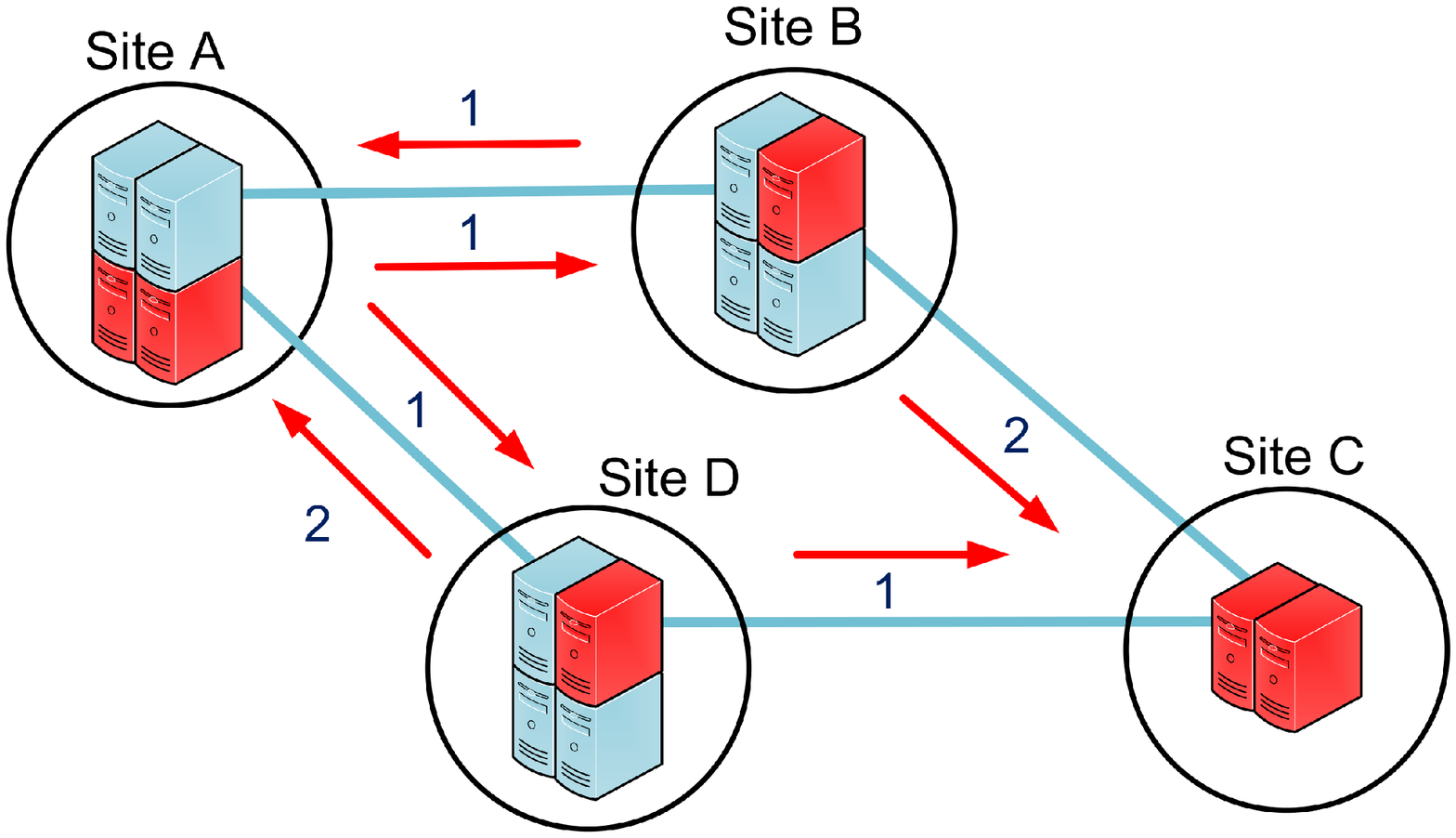}
\label{zeroRPOfig:modelA}}
\subfigure[Secunday path between sites A and D.]
{\includegraphics[width=0.45\textwidth]{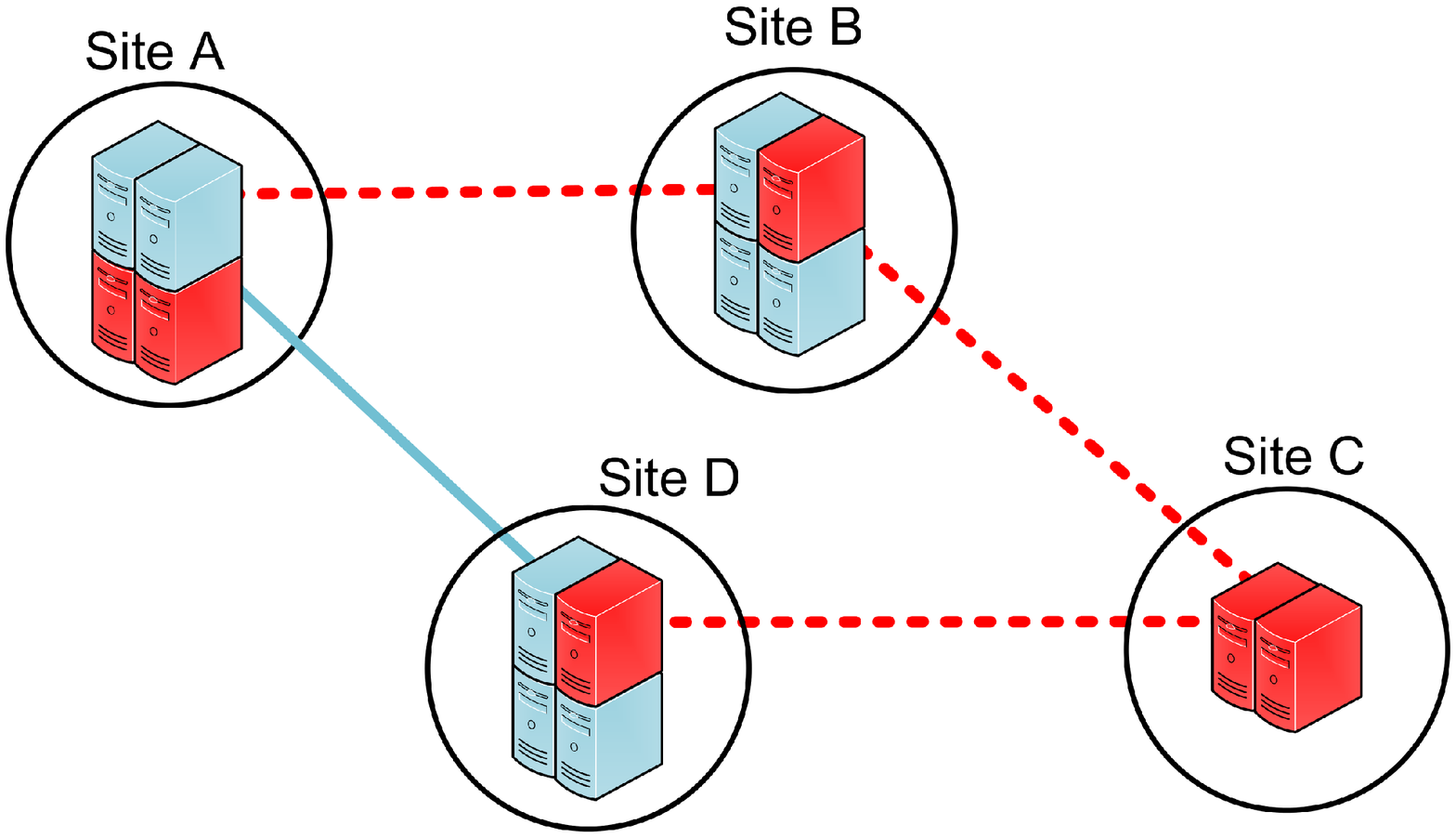}
\label{zeroRPOfig:modelB}}
\caption{Example of geo-distributed DC with continuous VM replication.}
\label{zeroRPOfig:model}
\end{figure}

Figure~\ref{zeroRPOfig:modelA} exemplifies the considered scenario, with a DC designed to support single-site failures. Each circle represents a site placed in a geographical location, and all sites are interconnected by a WAN. Note that the VMs hosted in each server are not shown in the figure. The arrows indicate that a site continuously sends VM state replicas to its neighbors.
The numbers next to each arrow indicate how many primary servers in the source site send backups to the destination site. For example, Site D has three primary servers, and sends VM backups of two servers to Site D, while the third backup is sent to Site C. The figure also shows that a single site can have both primary and backup servers. 

Figure~\ref{zeroRPOfig:modelA} also shows that a site can receive more backups of primary servers than the number of servers with this function. For example, Site C receives backups from two servers in Site B and one server in Site D, which would need three backup servers in Site C. However, Site C has only two backup servers due to the backup sharing scheme proposed in this work. 
Considering that sites B and D do not fail at the same time, Site C does not need to host operational VMs from three primary servers at the same time. As the service is based on virtualization, in a normal DC operation (i.e., with no failure) a backup server does not need to maintain operational its backup VMs, storing only data related do disk, memory content, and other VM information sent by the primary server~\cite{couto2014Network,wood2010disaster}. In normal operation, the backup server needs only VM storage capacity, provided by storage servers, not shown in the figure. The memory and CPU capacity, provided by the backup servers, is only needed after a disaster, when the recovery procedures are executed and the backup VMs start to run. We can thus reduce the number of backup servers, since a site needs only to support the worst-case failure of a single site.
In the case of Site C, the worst-case is the failure of Site B, which requires two of its servers to become operational.
This backup sharing scheme is considered in the proposed optimization problem, allowing a significant reduction on the number of backup servers. It is important to note that, despite the use of this scheme, the number of storage servers is always the same. Consequently, we do not consider the placement of this type of server.

A basic requirement of the server placement is that a primary server and its corresponding backup must be placed in different sites, and must not fail at the same time. To this end, we use the Failure Independence Matrix (matrix $I$).
Each element in $I$ has a binary value $I_{ij}$, which is 1 if site $i$ can become unreachable at the same time as site $j$, and 0 otherwise. A site is considered unreachable if, after a disaster, it does not have a path to a gateway or if the site itself is down. The matrix $I$ is built using a failure model. In this work, we consider the single-failure model, detailed later in this article.

\subsection{Replication Link and Secondary Path}

As the VM replication needs a very low latency between primary and backup servers, we force a given site to replicate backups only in its one-hop neighbors. Hence, we avoid the latency increase caused by transmission and processing delays in routers along the path. We thus use the link between the primary server and its corresponding backup, called here the replication link, to perform VM replication. If this link fails and the two sites cannot communicate with each other, the replication processes stops and the VMs of the primary server start to run in the unprotected mode~\cite{rajagopalan2012secondsite}.
In this mode, the VMs are still operational but are not replicated, since the communication with the backup server is broken.
As in services with zero RPO the unprotected mode should be avoided, in this work we configure secondary paths between the primary server and its backup. Consequently, when the replication link is broken, the primary server sends the VM replication through the secondary path. Obviously, this path must not contain the replication link.
Figure~\ref{zeroRPOfig:modelB} shows the secondary path between sites A and D. Later in this article, we analyze the tradeoffs of using secondary paths, as well as the quality of these paths.

\section{Server Placement Problem Formulation}
\label{zeroRPOsec:problem}

In this work, we maximize the number of primary servers covered by the continuous backup service and jointly minimize the number of backup servers installed. The optimization problem takes as parameters the link latency (in ms) and capacity (in Mbps)\footnote{Although the link capacities in the considered networks are generally in the order of Gbps, we use Mbps since it is suitable to the values of bandwidth consumption employed in our evaluation, as seen later.}, the Failure Independence Matrix, as well as the network topology to evaluate the secondary paths. The problem output provides the number of primary and backup servers installed in each site, as well as the secondary paths between each pair of primary and backup sites. The proposed placement scheme is performed in two steps. The first one evaluates the secondary paths between each pair of sites in the network. The second step executes the physical server optimization problem.

In the first step, we model the WAN topology as a graph, in which vertices are sites and each edge is a link. The weight of each link is its latency, which is directly proportional to the geographical distance between the two sites connected by this link. To evaluate the secondary paths for each pair of one-hop neighbor sites, we remove from the graph the link between these two sites. Then, we reevaluate the shortest path between these sites using the Dijkstra algorithm. Using this strategy, we choose the secondary path with the lowest possible latency, regardless of the optimization problem employed in the next step. We adopt this strategy to force low latency values due to the strict requirement regarding this metric. It is worth noting that, at the end of the next step, we do not configure secondary paths between sites that do not replicate between each other.
The result of this first step is the set of $s^{km}_{ij}$ parameters (Table~\ref{zeroRPOtab:notations}), which defines the links belonging to the secondary path between each pair of sites $k$ and $m$. Hence, for each link between $i$ and $j$, we have $s^{km}_{ij} = 1$ if this link appears in the secondary path between $k$ and $m$, and $s^{km}_{ij} = 0$, otherwise.

The second step solves the ILP (Integer Linear Programming) problem formulated hereinafter. This problem chooses the placement of each primary server and its corresponding backup, considering the aforementioned optimization goals and restrictions. Table~\ref{zeroRPOtab:notations} lists the main notations used in this work, as well as the type of each one. Notations with \textit{set} or \textit{parameter} types are the problem input, while the \textit{variables} are adjusted by the optimization algorithm. The ILP problem is formulated as follows:
\begin{table}
\caption{Notations used in the problem formulation.}
\centering
\scriptsize
\begin{center}
\begin{tabular}{|c||l||c|}
\hline \textbf{Notation} &\textbf{Description} &\textbf{Type}\\
\hline $\mathcal{D}$ &Candidate Sites &Set\\
\hline \multirow{2}{*}{$I_{ij}$} &Binary value indicating whether site $i$ can become unreachable &\multirow{2}{*}{Parameter}\\
&at the same time as site $j$  &\\
\hline $\Delta_{ij}$ &Propagation delay (latency) of the link between sites $i$ and $j$ &Parameter\\
\hline $W_{ij}$ &Link capacity between sites $i$ and $j$ &Parameter\\
\hline \multirow{2}{*}{$s^{km}_{ij}$} &Binary value indicating whether the link between $i$ and $j$ belongs &\multirow{2}{*}{Parameter}\\
&to the secondary path between $k$ and $m$ &\\
\hline $\alpha$ &Maximum fraction of the link capacity allowed to the replication &Parameter\\
\hline \multirow{2}{*}{$\gamma$} &Binary value indicating if secondary paths will &\multirow{2}{*}{Parameter}\\
&be deployed in the WAN infrastructure&\\
\hline \multirow{2}{*}{$B$} &Bandwidth consumption, in Mbps, as a consequence of the  &\multirow{2}{*}{Parameter}\\
&continuous VM replication &\\
\hline \multirow{2}{*}{$L_{worst}$} &Maximum latency allowed between a primary server and  &\multirow{2}{*}{Parameter}\\
&its corresponding backup &\\
\hline \multirow{2}{*}{$U_{max}$} &Maximum number of active sites (i.e., sites with at least   &\multirow{2}{*}{Parameter}\\
&one installed server)&\\
\hline $x_i$ & Number of primary servers in location $i$ &Variable\\
\hline $b_i$  & Number of backup servers in location $i$ &Variable\\
\hline $u_i$  &Binary value indicating whether site $i$ is active ($(x_i + b_i) > 0$) &Variable\\
\hline $r_{ij}$ &Amount of bandwidth available to replicate from site $i$ to site $j$ &Variable\\
\hline $c_{ij}$ &Number of primary server backups that site $i$ sends to site $j$ &Variable\\
\hline \multirow{2}{*}{$e_{ij}$} &Binary value indicating whether site $i$ replicates backups &\multirow{2}{*}{Variable}\\
&to site $j$ ($c_{ij} > 0$) &\\
\hline \multirow{2}{*}{$y_{kij}$} &Binary value indicating whether site $k$ receives backups &\multirow{2}{*}{Variable}\\
&from sites $i$ and $j$ ($e_{ik} = 1$ and $e_{jk} = 1$) &\\
\hline
\end{tabular}
\label{zeroRPOtab:notations}
\end{center}
\end{table}
 
\begin{equation}
\textup{maximize}\ \  \sum_{i \in \mathcal{D}} (x_i - b_i)
\label{zeroRPOlp:objective}
\end{equation}

\begin{equation}
\textup{subject  to} \ \ \ c_{ij} \ I_{ij} = 0 \quad \forall i,j \in \mathcal{D}
\label{zeroRPOlp:backupDifferentSRG}
\end{equation}
\begin{equation}
\sum_{j \in \mathcal{D}} c_{ij} = x_i \quad \forall i \in \mathcal{D}
\label{zeroRPOlp:totalBackupsSite}
\end{equation}
\begin{equation}
M \ e_{ij} - c_{ij} \geq 0 \quad \forall i,j \in \mathcal{D}
\label{zeroRPOlp:evaluateEijA}
\end{equation}
\begin{equation}
e_{ij} \leq c_{ij} \quad \forall i,j \in \mathcal{D}
\label{zeroRPOlp:evaluateEijB}
\end{equation}
\begin{equation}
y_{kij} \geq e_{ik} + e_{jk} - 1 \quad \forall k,i,j \in \mathcal{D}, \ i < j
\label{zeroRPOlp:evaluateEijC}
\end{equation}
\begin{equation}
y_{kij} \leq e_{ik} \quad \forall k,i,j \in \mathcal{D}, \ i < j
\label{zeroRPOlp:evaluateEijD}
\end{equation}
\begin{equation}
y_{kij} \leq e_{jk} \quad \forall k,i,j \in \mathcal{D}, \ i < j
\label{zeroRPOlp:evaluateEijE}
\end{equation}
\begin{equation}
\sum_{i,j \in \mathcal{D}, \ i < j} (I_{ij} \ y_{kij}) = 0 \quad \forall k \in \mathcal{D}
\label{zeroRPOlp:savePhysicalServers}
\end{equation}
\begin{equation}
b_j - c_{ij} \geq 0 \quad \forall i,j \in \mathcal{D}
\label{zeroRPOlp:maxDedicatedServers}
\end{equation}
\begin{equation}
B \ c_{ij} \leq r_{ij} \quad \forall i,j \in \mathcal{D}
\label{zeroRPOlp:bandwidthDemand}
\end{equation}
\begin{equation}
r_{ij} \leq \alpha \ W_{ij} - \gamma \ B \ c_{km} \ s^{km}_{ij} \quad \forall i,j \in \mathcal{D} \quad \forall k,m \in \mathcal{D}
\label{zeroRPOlp:backupPath}
\end{equation}
\begin{equation}
e_{ij} \ \Delta_{ij} \leq L_{worst} \quad \forall i,j \in \mathcal{D}
\label{zeroRPOlp:latencyDemand}
\end{equation}
\begin{equation}
M \ u_i - (x_i + b_i) \geq 0 \quad \forall i \in \mathcal{D}
\label{zeroRPOlp:evaluateUiA}
\end{equation}
\begin{equation}
u_i \leq (x_i + b_i) \quad \forall i \in \mathcal{D}
\label{zeroRPOlp:evaluateUiB}
\end{equation}
\begin{equation}
\sum_{i \in \mathcal{D}} u_i \leq U_{max}
\label{zeroRPOlp:maxNrSites}
\end{equation}
\begin{equation}
x_i \geq 0, \quad b_i \geq 0, \ \forall \ i \in \mathcal{D} ; c_{ij} \geq 0 \ \forall \ i,j \in \mathcal{D}
\label{zeroRPOlp:boundsA}
\end{equation}
\begin{equation}
\begin{split}
\quad x_i \in \mathbb{Z}, \quad b_i \in \mathbb{Z} \ \forall \ i \in \mathcal{D}; \quad r_{ij} \in \mathbb{Z}, \ c_{ij} \in \mathbb{Z} \ \forall \ i,j \in \mathcal{D}; u_i \in \{0,1\},  \ \forall \ i \in \mathcal{D}; \\
 e_{ij} \in \{0,1\} \ \forall \ i,j \in \mathcal{D}; y_{kij} \in \{0,1\} \ \forall \ k,i,j \in \mathcal{D}
\end{split}
\label{zeroRPOlp:domain}
\end{equation}

The objective of this problem, given by Equation~\ref{zeroRPOlp:objective}, is to maximize the number of primary servers ($\sum_{i \in \mathcal{D}} x_i$) and to minimize the number of backup servers ($\sum_{i \in \mathcal{D}} b_i$). For each server installed on a site, the problem tries to reduce by one unit the number of backup servers. Hence, the objective function can be seen as the savings in the number of backup servers installed. In the worst case, the objective function is zero; whereas in the best case, it tends to be close to the number of primary servers installed.

Equation~\ref{zeroRPOlp:backupDifferentSRG} forces the primary servers of site $i$ to store their backups on site $j$ (i.e., $c_{ij} > 0$) only if $i$ and $j$ cannot become unreachable at the same time ($I_{ij} = 0$). 
Equation~\ref{zeroRPOlp:totalBackupsSite} defines that the number of server backups replicated by site $i$ must be equal to the number of primary servers installed in this site.

We use Equations~\ref{zeroRPOlp:evaluateEijA}~and~\ref{zeroRPOlp:evaluateEijB} to evaluate the binary variables $e_{ij}$, which receive 1 if $c_{ij} > 0$ and 0, otherwise. The parameter $M$ in Equation~\ref{zeroRPOlp:evaluateEijA} is just a high value, set to be always greater or equal to any possible value of the variables 
$c_{ij}$ in this equation, and the sum $x_i + u_i$ in Equation~\ref{zeroRPOlp:evaluateUiA}, shown later. We adopt, conservatively, $M=1\times10^9$.
The Equations~\ref{zeroRPOlp:evaluateEijC},~\ref{zeroRPOlp:evaluateEijD}~and~\ref{zeroRPOlp:evaluateEijE}, employed to evaluate the binary variables $y_{kij}$, are together equivalent to the logical operation AND between $e_{ik}$ and $e_{jk}$.

Equation~\ref{zeroRPOlp:savePhysicalServers} is responsible for saving backup servers. This equation ensures that if two sites $i$ and $j$ can become unreachable at the same time (i.e., $I_{ij} = 1$), they cannot host their backup in the same site $k$. In other words, if $I_{ij} = 1$, $i$ and $j$ cannot share backup resources. Note that Equation~\ref{zeroRPOlp:savePhysicalServers} affects the variables $y_{kij}$, and thus the variables $e_{ij}$ and $c_{ij}$, which define the backup placement.
As the problem ensures that sites that can become unreachable at the same time do not share backup sites, the number of backup servers can be evaluated by using 
Equation~\ref{zeroRPOlp:maxDedicatedServers}, which is equivalent to $b_j = \max_{j \in \mathcal{D}} (c_{ij})$. Hence, the number of backup servers installed in a site $j$ is given by the maximum number of backups that any other site in the network sends to it.
Using again the example of Figure~\ref{zeroRPOfig:modelA}, Site C receives two backups from Site B and one backup from Site D. Consequently, Site C has two backup servers to support the failure of Site B, which assigns to Site C the highest number of primary server replications.

Equations~\ref{zeroRPOlp:bandwidthDemand}~and~\ref{zeroRPOlp:backupPath} take care of the bandwidth restrictions. In Equation~\ref{zeroRPOlp:bandwidthDemand}, $B \ c_{ij}$ corresponds to the amount of bandwidth required to replicate $c_{ij}$ primary servers from $i$ to $j$.
Note that, for each replication, $B$ Mbps are continuously used in the link between the two sites. Equation~\ref{zeroRPOlp:bandwidthDemand} specifies that the bandwidth consumption must be smaller or equal than the value defined by $r_{ij}$, which is given by Equation~\ref{zeroRPOlp:backupPath}.
This equation defines the amount of bandwidth available $r_{ij}$ to replicate between sites $i$ and $j$, considering also the bandwidth reserved by the secondary paths that contains the link between $i$ and $j$. In Equation~\ref{zeroRPOlp:backupPath}, $\alpha \ W_{ij}$ is the total bandwidth available in each link to the disaster-resilient IaaS service. The parameter $W_{ij}$ is the link capacity, which is zero if sites $i$ and $j$ do not have a link between each other. Consequently, a given site can only replicate backups to its one-hop neighbors. The term $ \ \gamma \ B \ c_{km} \ s^{km}_{ij} \ $ is the amount of bandwidth used by a given secondary path between two sites $k$ and $m$, that contains the link between $i$ and $j$. 
Note that, in this problem, Equation~\ref{zeroRPOlp:backupPath} is equivalent to $r_{ij} = \min_{k,m \in \mathcal{D}} (\alpha \ W_{ij} - B  \ c_{km} \ s^{km}_{ij})$. That is, considering all secondary paths that contain a given link, we account in $r_{ij}$ only the path which consumes the highest bandwidth amount.
This is possible since we assume that \textit{the links in the network do not fail simultaneously}, and thus only one secondary path can be active in the whole DC. 
Given that, regarding the secondary paths, we only need to reserve in each link the amount of bandwidth required to the worst-case link failure.
In other words, the secondary paths are provisioned using a shared path protection scheme~\cite{ramamurthy2003survivable}.
The $\gamma$ parameter in Equation~\ref{zeroRPOlp:backupPath} is employed to disable secondary paths in the placement, freeing link resources that can be used to support more replications and thus more primary servers. Hence, if $\gamma = 0$, we have $r_{ij} = \alpha \ W_{ij}$.

Equation~\ref{zeroRPOlp:latencyDemand} defines the latency requirements. Hence, a given site $i$ only replicates to site $j$ (i.e., $e_{ij}=1$) if their replication link has a latency $\Delta_{ij}$ smaller or equal than the maximum latency allowed ($L_{worst})$. Equations \ref{zeroRPOlp:evaluateUiA}, \ref{zeroRPOlp:evaluateUiB}, and \ref{zeroRPOlp:maxNrSites} limit the number of sites chosen to install primary or backup servers, based on the maximum number of sites $U_{max}$. Finally, Equations~\ref{zeroRPOlp:boundsA}~and~\ref{zeroRPOlp:domain} define, respectively, the lower bound and the domain of each variable.  

In this work, we do not consider some common IaaS requirements, such as the latency between the end users and the VMs.
We disregard these requirements to allow a more detailed analysis of a zero RPO service, in which the main requirements are failure independence between primary servers and their corresponding backups, as well as the latency between them.
As we have already stated, the latency between backup and primary servers is a very important concern, since it affects directly the response time of VM applications.

\section{Evaluation}
\label{zeroRPOsec:eval}

\begin{figure}[ht!]
\centering
\subfigure[RNP, 28 sites and 38 links.]
{\includegraphics[width=0.48\textwidth]{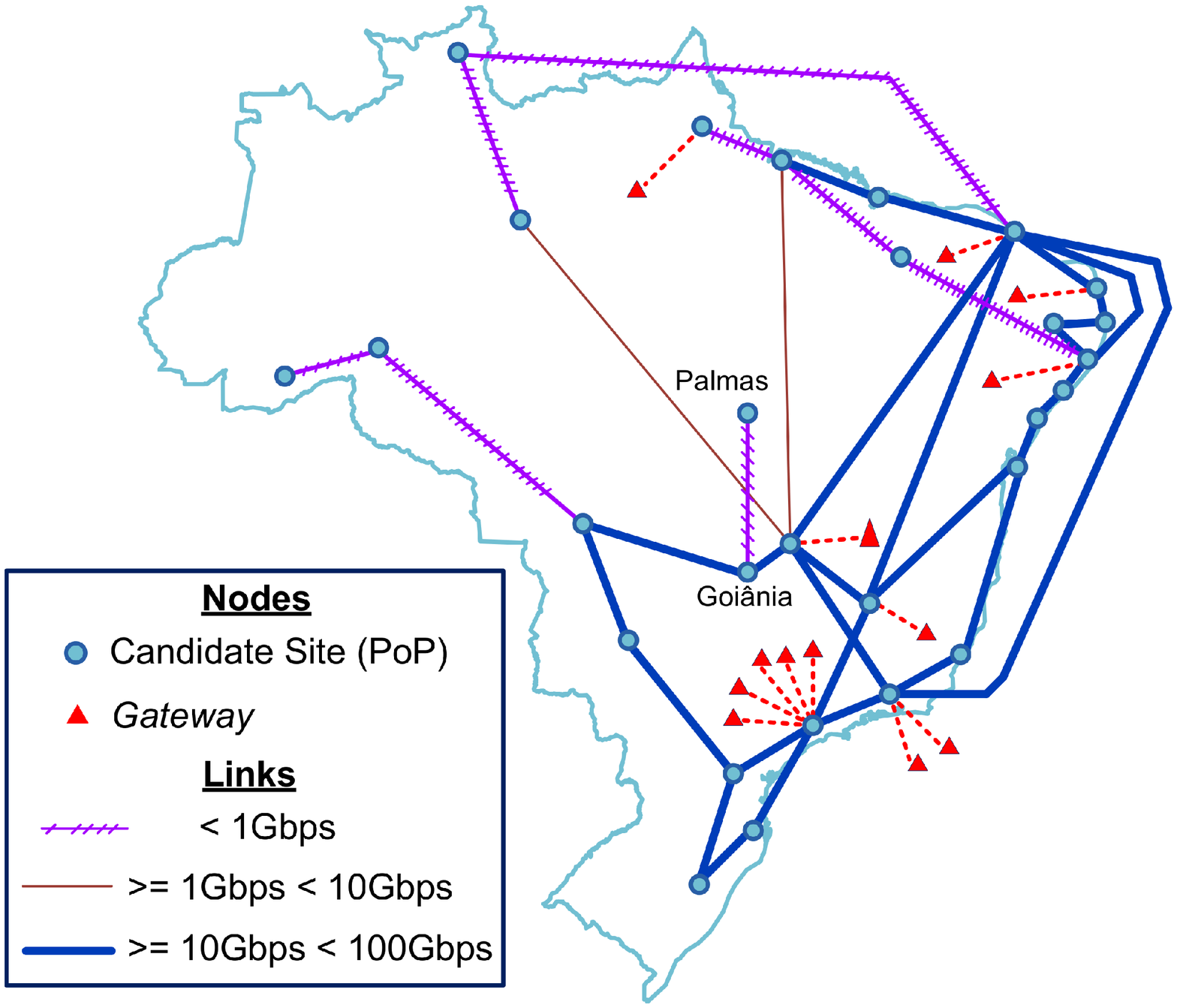}
\label{zeroRPOfig:redeIpe}}
\subfigure[RENATER , 48 sites and 67 links.]
{\includegraphics[width=0.48\textwidth]{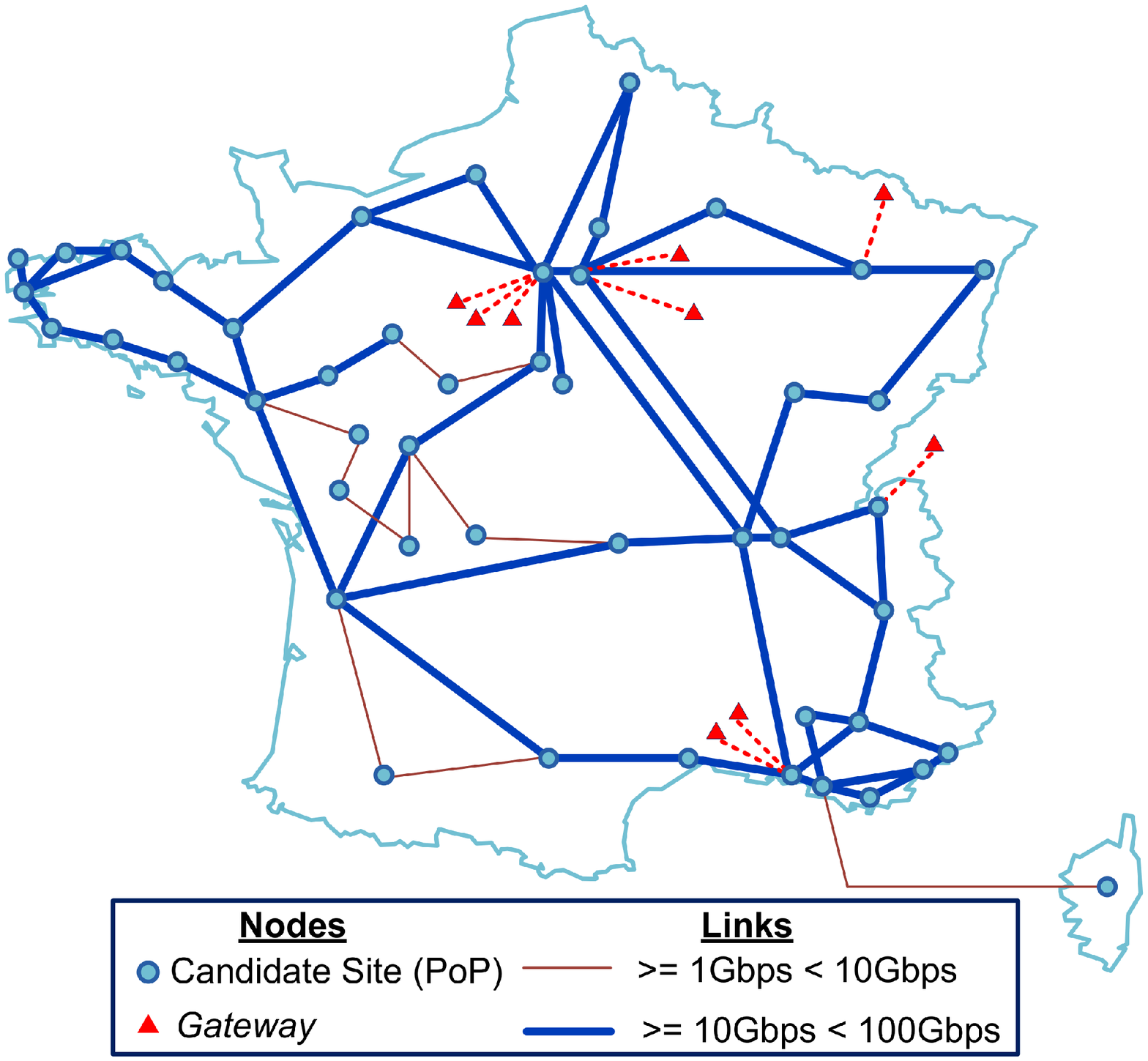}
\label{zeroRPOfig:renater}}
\subfigure[GEANT, 41 sites and 58 links.]
{\includegraphics[width=0.52\textwidth]{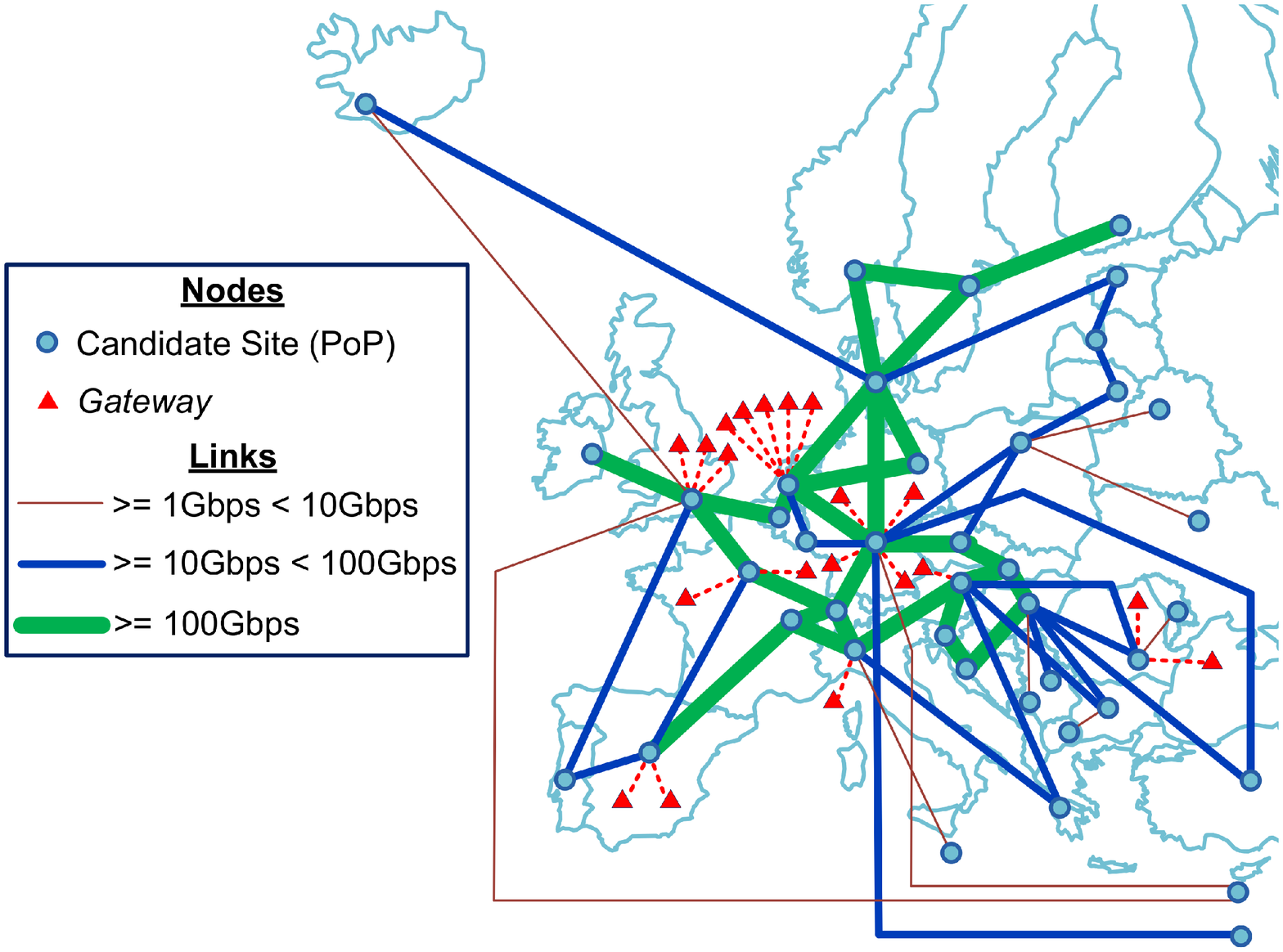}
\label{zeroRPOfig:geant}}
\caption{Topologies of the WANs considered in this work.}
\label{zeroRPOfig:topologies}
\end{figure}

The optimization problem formulated in Section~\ref{zeroRPOsec:problem} is employed in this work to place servers in real REN (Research and Educational Network) topologies. These WANs are composed of PoPs (Point of Presence) which, in the context of this work, are the candidate DC sites. We thus adopt the topologies from the Brazilian RNP (Figure~\ref{zeroRPOfig:redeIpe}), the French RENATER (Figure~\ref{zeroRPOfig:renater}), and the European GEANT (Figure~\ref{zeroRPOfig:geant}). Each subfigure of Figure~\ref{zeroRPOfig:topologies} shows, for each WAN, the DC sites, the gateways, and the link capacities.

From the mentioned topologies, we evaluate the input parameters of the optimization problem.
The latency value $\Delta_{ij}$ of each link is given by the propagation delay between sites $i$ and $j$.
We thus consider that the network is well provisioned and thus the queuing and transmission delays are negligible.
The propagation delay is directly proportional to the distance between the two sites, which is estimated in this work as the length of a straight line between the center of the two cities where the sites are installed. To evaluate this delay, we use a propagation speed of $2 \times 10^8$~m/s, which is commonly used in optical networks~\cite{couto2014Latency}. The Failure Independence Matrix (matrix $I$) is evaluated based on a single-failure model, employed also in our previous work~\cite{couto2014Latency}. Using this model, we consider that there are no simultaneous failures, which means that only one link or one DC site fails in a given instant. Note that, despite the single-failure model, two sites can become unreachable at the same time. For example, in the network of Figure~\ref{zeroRPOfig:redeIpe}, if the site in Goi\^ania is down, then the site in Palmas becomes unreachable. The capacity $W_{ij}$ of each link is shown in Figure~\ref{zeroRPOfig:topologies}. These are real values, extracted from the website of each REN. The network is modeled as a directed graph, which means that the capacities shown in the figure are the bandwidth values on each link direction. Although the graph is directed, the failure model considers that if a given link is down, then both directions are down.

The bandwidth consumption, generated by the continuous replication of each primary server, is given by the $B$ parameter, fixed at 240 Mbps. 
This value is extracted from the SecondSite paper~\cite{rajagopalan2012secondsite}, and corresponds approximately to the bandwidth consumption when replicating a primary server with four VMs, two running a web server benchmark and two running a database benchmark. 
The value for $B$ chosen in this work is employed only as a reference, since it can vary significantly depending on the applications running on the VMs as well as the load imposed by users. Hence, when designing a disaster-resilient DC, $B$ must be chosen according to the SLAs (Service Level Agreements), and thus the provider must reserve a given bandwidth to the replication service. Furthermore, given the heterogeneity of the applications running in an IaaS infrastructure, we can have in a single DC different values for $B$. We use in this work only one value for $B$ to simplify our analysis. However, our problem can be easily modified to consider different $B$ parameters. Another parameter fixed for all evaluations in this section is the maximum number of active sites, given by $U_{max}$. In this analysis, we choose a very high value for this parameter, so as not to limit the number of used sites. Finally, if not mentioned otherwise, the $\gamma$ parameter is fixed at 1. This means that the problem considers the configuration of secondary paths.

We use the graph manipulation tool NetworkX~\footnote{The NetworkX tool is available in \url{http://networkx.github.io/}} to generate the Failure Independence Matrix and to run the first optimization step. The ILP (Integer Linear Programming) problem, corresponding to the second optimization step, is solved using IBM ILOG CPLEX~\footnote{Details about CPLEX are available in \url{http://www-01.ibm.com/software/commerce/optimization/cplex-optimizer/}} 12.5.1.

\subsection{Service Capacity and Savings}

\begin{figure}[ht!]
\centering
\subfigure[RNP.]
{\includegraphics[width=0.40\textwidth]{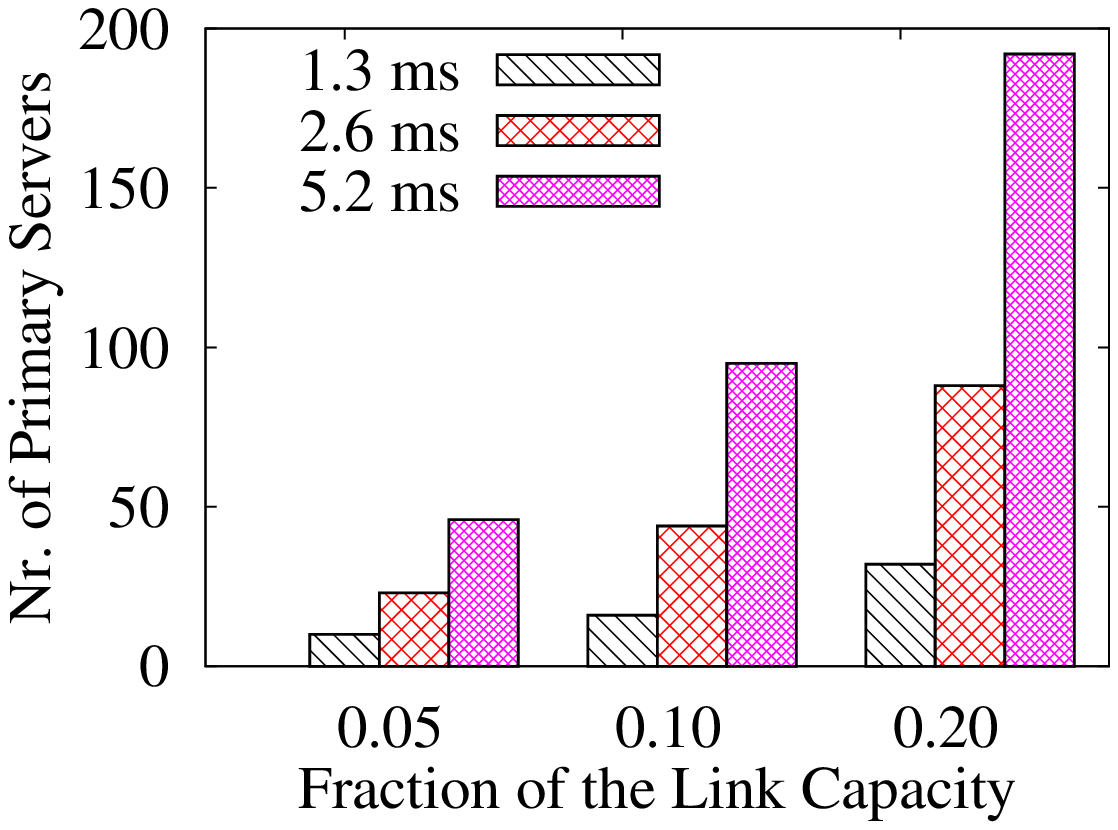}
\label{zeroRPOfig:totalNrServers_rnp}}
\subfigure[RENATER.]
{\includegraphics[width=0.40\textwidth]{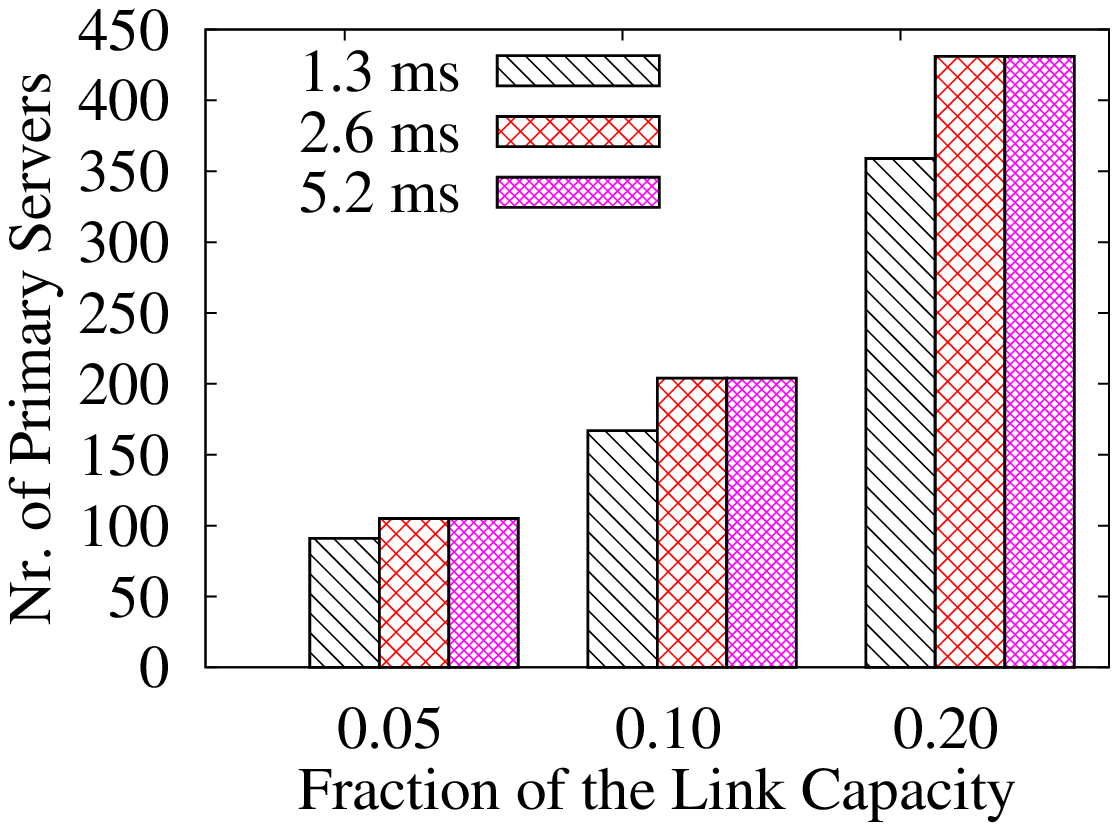}
\label{zeroRPOfig:totalNrServers_renater}}
\subfigure[GEANT.]
{\includegraphics[width=0.40\textwidth]{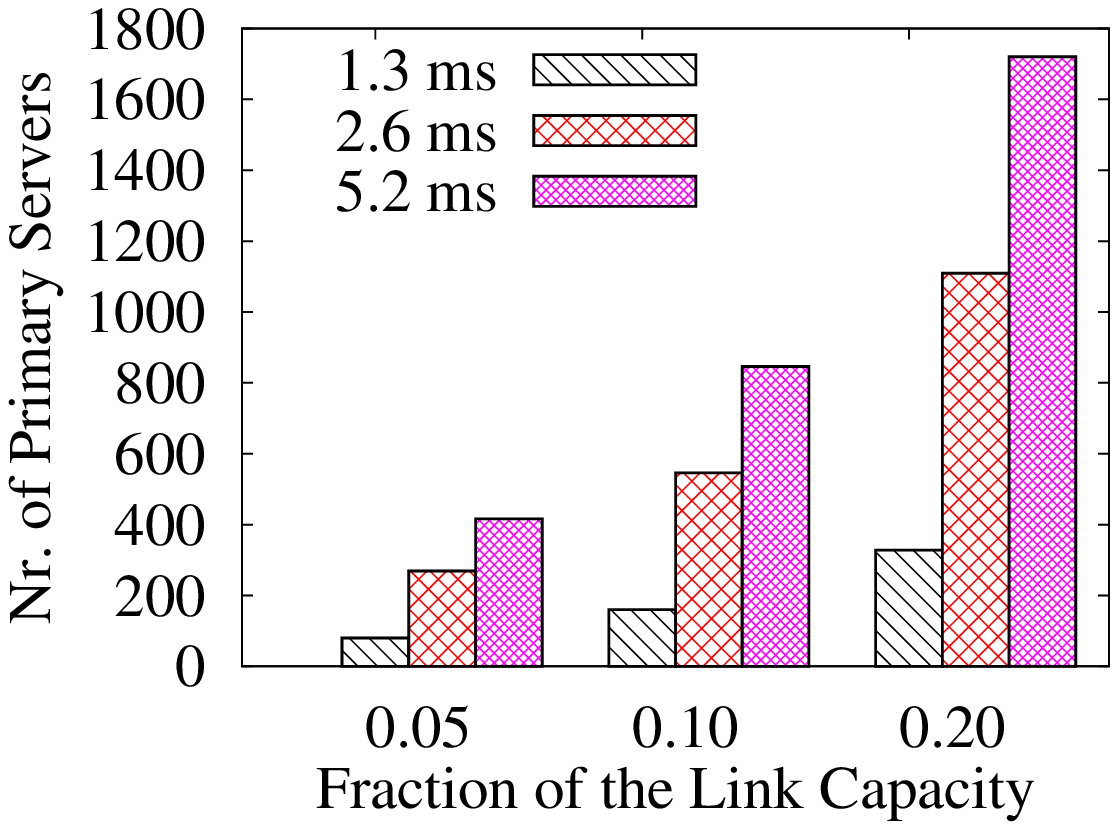}
\label{zeroRPOfig:totalNrServers_geant}}
\caption{IaaS service Capacity.}
\label{zeroRPOfig:totalNrServers}
\end{figure}

We run our optimization problem with the chosen parameters, and we analyze the service capacity of the disaster-resilient cloud deployed in each one of the three considered networks. The IaaS service capacity, given by the number of primary servers supported, is evaluated for different values of allowed bandwidth fraction ($\alpha$) and different values of maximum tolerated latency ($L_{worst}$). 
Figure~\ref{zeroRPOfig:totalNrServers} shows the service capacity results. For each network, the plot groups in the X-axis results for different $\alpha$ values, while each similar bar corresponds to a different $L_{worst}$. The lowest $L_{worst}$ (i.e., 1.3~ms) is chosen according to experiments conducted in the SecondSite article~\cite{rajagopalan2012secondsite}. These experiments consist of replicating a server on a 260~km link between the Canadian cities of Vancouver and Kamloops. This distance implies a propagation delay of 1.3~ms, considering the propagation speed employed in our work. The other two $L_{worst}$ values used in this work are relaxation of the latency requirements, being the double and the quadruple of the reference value of 1.3~ms.
Results in Figure~\ref{zeroRPOfig:totalNrServers} show, as expected, that we increase the number of primary servers when we increase the bandwidth fraction or when we relax latency requirement. Note that in RENATER the number of primary servers remains the same when we relax the latency requirement from 2.6~ms to 5.2~ms. This happens because RENATER is located in the metropolitan France, which is an area significantly smaller than the area spanned by RNP and GEANT. Consequently, the majority of RENATER links already meets the latency requirement when $L_{worst}=2.6$~ms.

\begin{figure}[ht!]
\centering
\subfigure[RNP.]
{\includegraphics[width=0.40\textwidth]{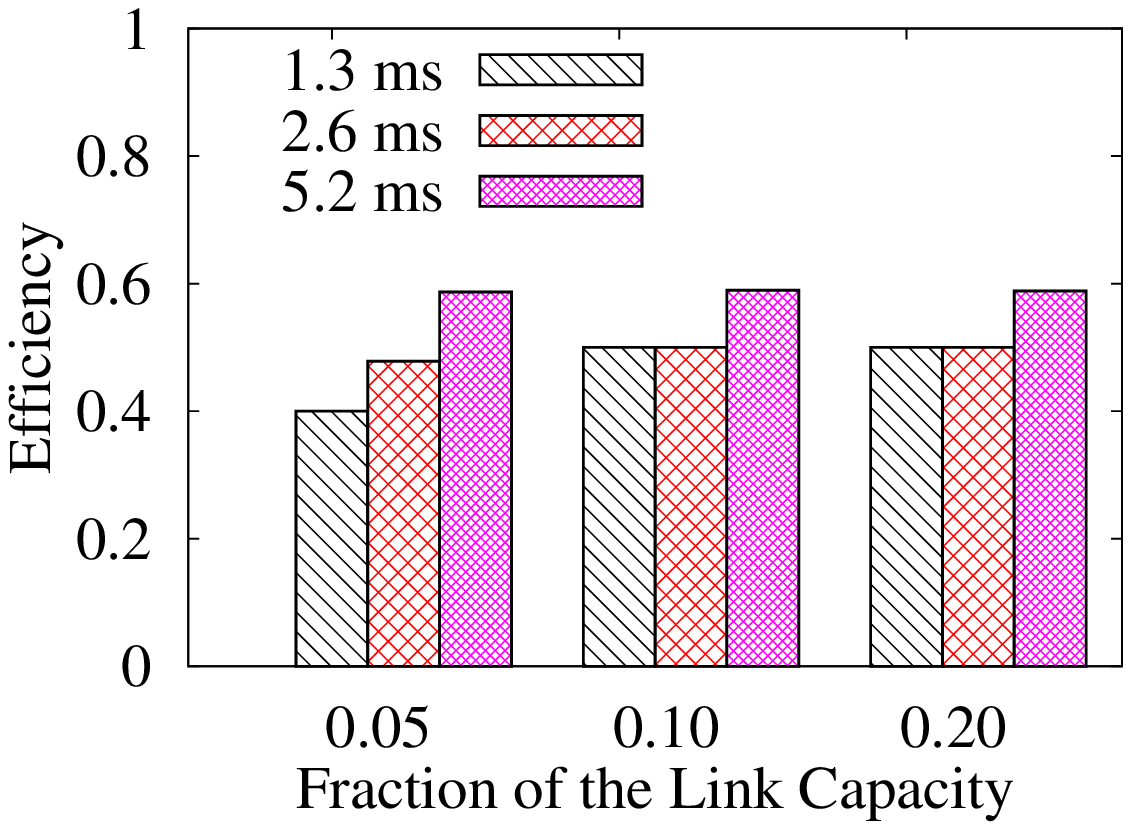}
\label{zeroRPOfig:ratioIaaSservers_rnp}}
\subfigure[RENATER.]
{\includegraphics[width=0.40\textwidth]{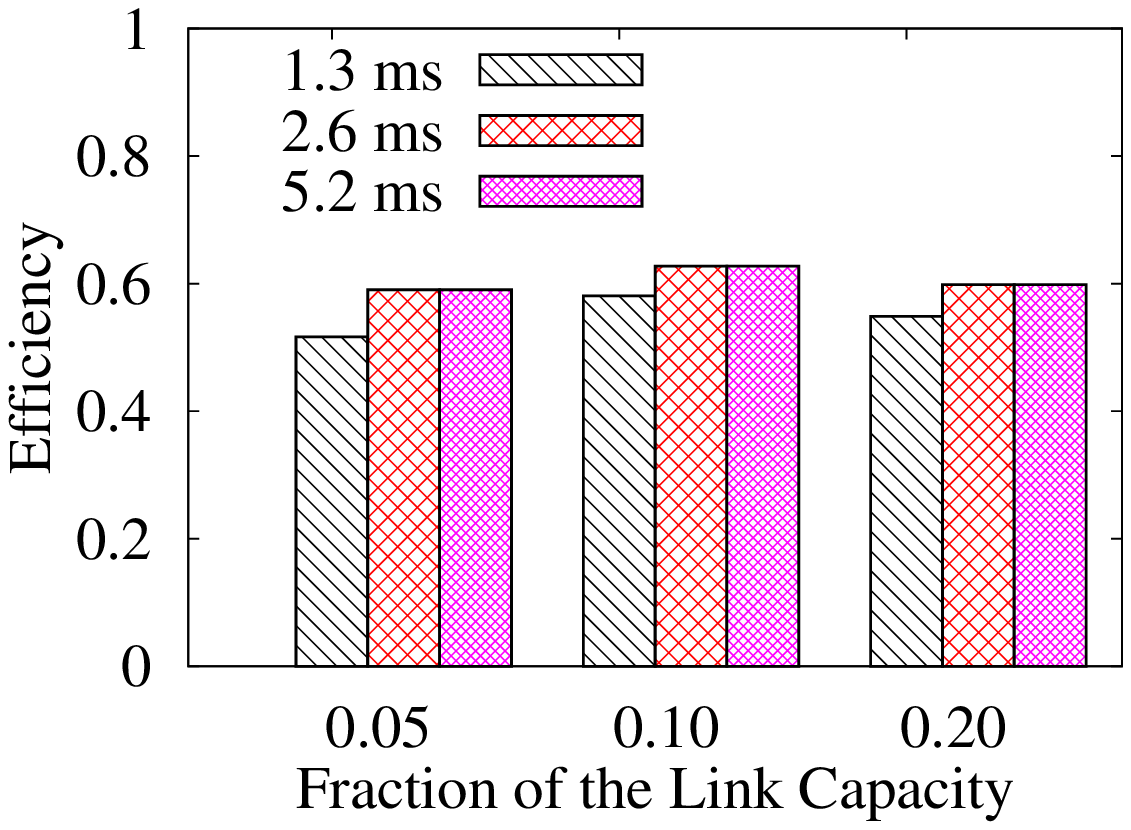}
\label{zeroRPOfig:ratioIaaSservers_renater}}
\subfigure[GEANT.]
{\includegraphics[width=0.40\textwidth]{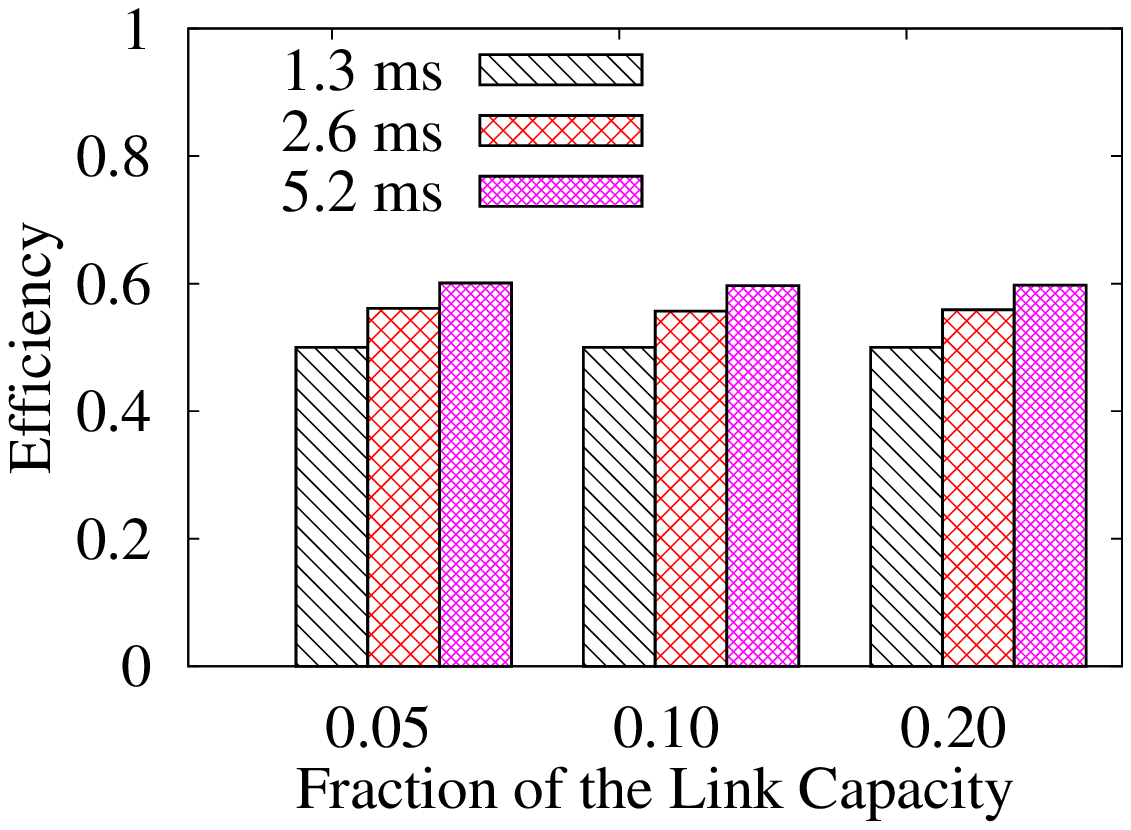}
\label{zeroRPOfig:ratioIaaSservers_geant}}
\caption{Backup server savings.}
\label{zeroRPOfig:ratioIaaSservers}
\end{figure}

Figure~\ref{zeroRPOfig:ratioIaaSservers} shows the savings on backup servers achieved by the placement optimization. The savings are quantified by the metric Server Efficiency (SE), defined as the relationship between the reduction of backup servers provided by our scheme and the total number of primary servers, given by:
\begin{equation}
\text{SE} = \frac{ \sum_{i \in \mathcal{D}} (x_i  - b_i)}{\sum_{i \in \mathcal{D}} (x_i)} = 1 - \frac{\sum_{i \in \mathcal{D}} (b_i)}{\sum_{i \in \mathcal{D}} (x_i)}.
\end{equation}

According to the definition, the SE metric lies in the interval $[0,1[$, and the higher its value, the higher the savings. The classical scheme, where no backup servers are shared, i.e., $\sum_{i \in \mathcal{D}} (b_i)=\sum_{i \in \mathcal{D}} (x_i)$, present a zero efficiency. The highest efficiency value is given by the placement where only one backup server is shared by all primary servers, i.e, $\sum_{i \in \mathcal{D}} (b_i)=1$. 
Figure~\ref{zeroRPOfig:ratioIaaSservers} shows that, for all networks, the efficiency is equal to or greater than 40\%, considering the $\alpha$ and $L_{worst}$ values evaluated. This shows that the proposed placement allow significant savings in the number of backup servers. Furthermore, the results show that relaxing the latency requirement can improve the efficiency, since it gives the problem more options to perform the placement. 

\subsection{Secondary Paths}
\label{zeroRPOsubsec:caminhosSecundarios}

As stated before, the placement in this work limits the maximum latency value in the replication link between two sites. However, the optimization does not limit the latency of the secondary paths. This means that, if the replication link between two sites fails, they need to replicate their VMs using a path that may not meet the latency requirement defined by $L_{worst}$. Hence, the response time of VM applications can increase. A na\"{\i}ve solution to this shortcoming is to place primary servers and their backups only in pair of sites which have secondary paths meeting the $L_{worst}$. However, as we show later in this paper, the number of pairs with secondary paths meeting this requirement is very low. Using thus only these site pairs would reduce significantly the number of primary servers supported.

\begin{figure}[ht!]
\centering
\subfigure[RNP]
{\includegraphics[width=0.42\textwidth]{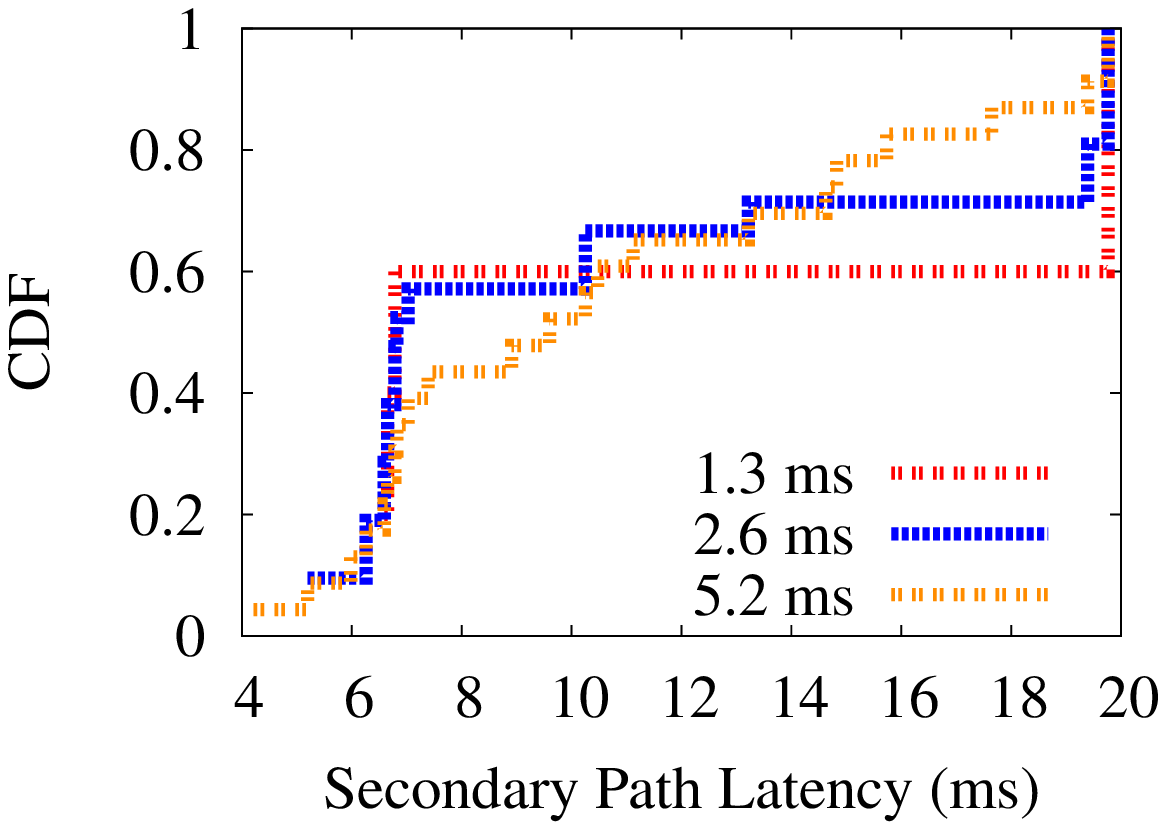}
\label{zeroRPOfig:cdfRedundantPathLatencyRNP}}
\subfigure[RENATER]
{\includegraphics[width=0.42\textwidth]{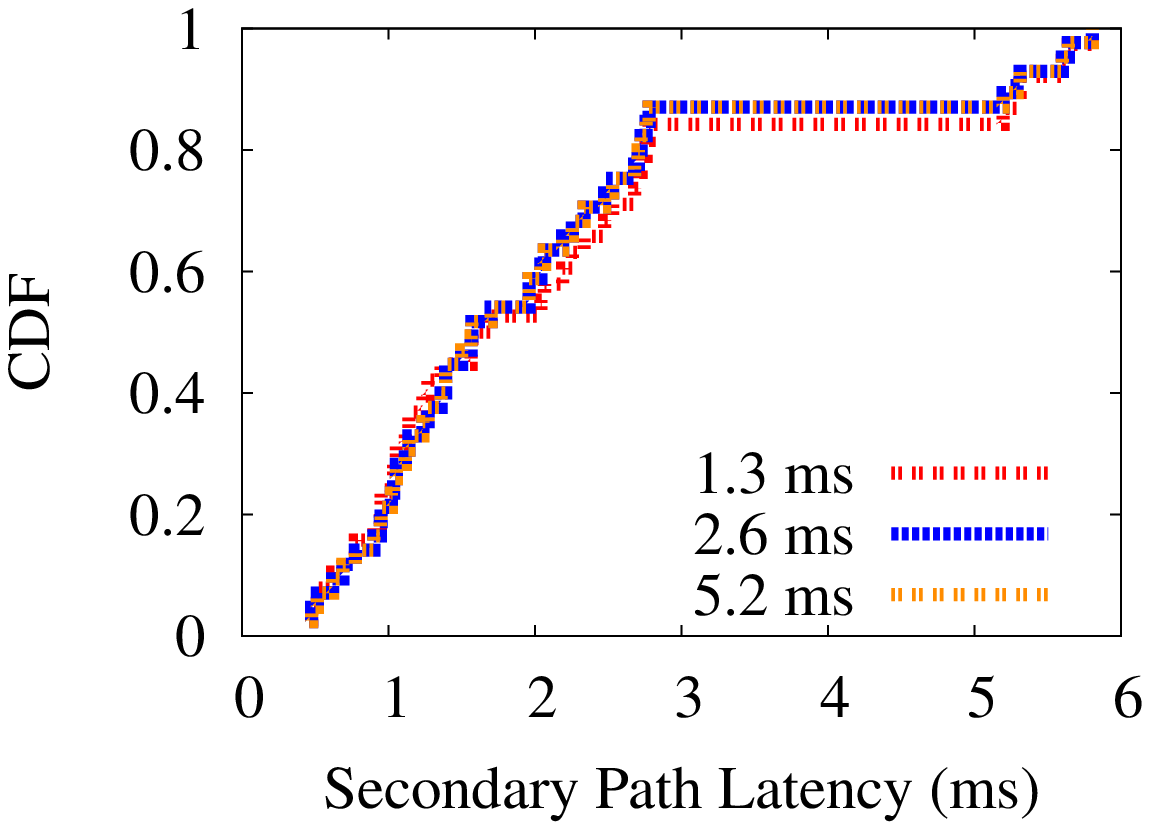}
\label{zeroRPOfig:cdfRedundantPathLatencyRENATER}}
\subfigure[GEANT]
{\includegraphics[width=0.42\textwidth]{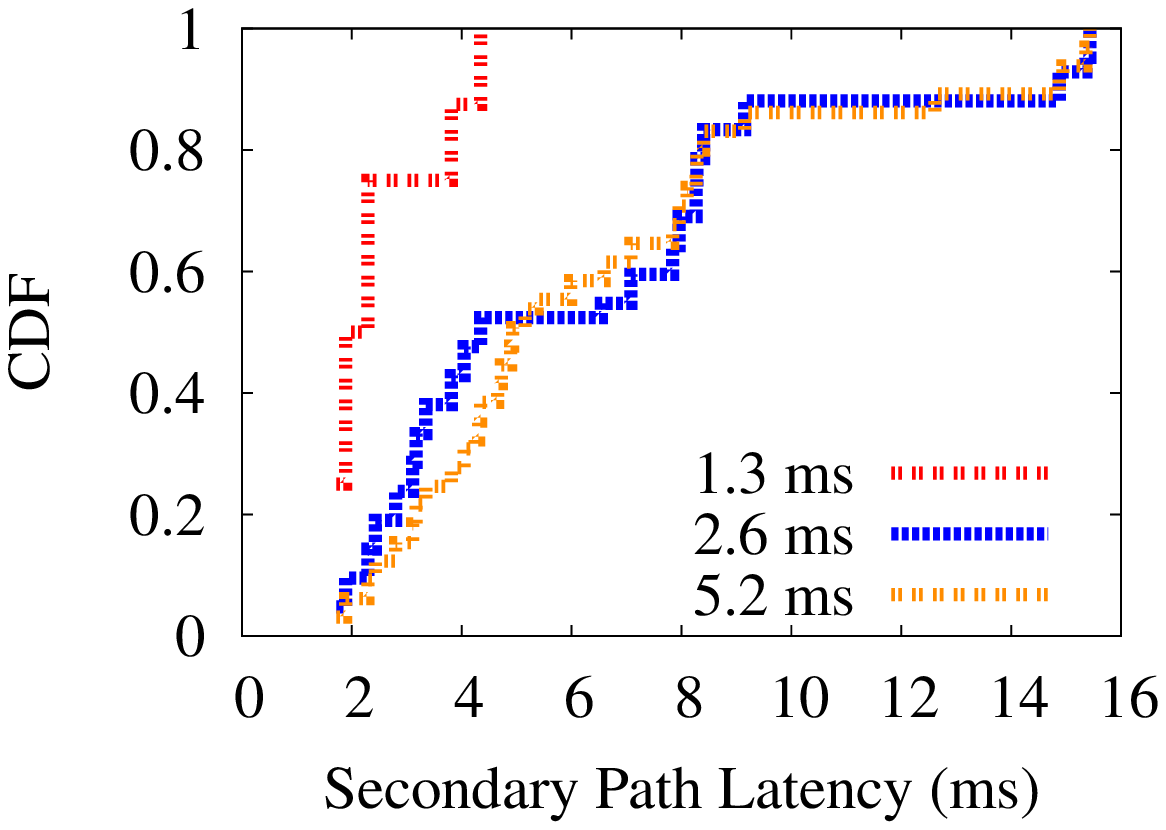}
\label{zeroRPOfig:cdfRedundantPathLatencyGEANT}}
\caption{Latency CDF of the secondary paths ($\alpha = 0.05$).}
\label{zeroRPOfig:redudantPaths}
\end{figure}

\begin{figure}[ht!]
\centering
\includegraphics[width=0.40\textwidth]{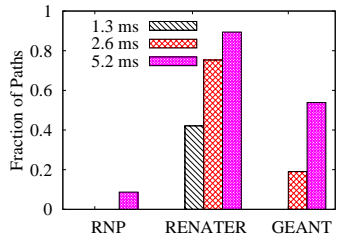}
\caption{Fraction of paths meeting $L_{worst}$.}
\label{zeroRPOfig:percentComplyPaths}
\end{figure}

Figure~\ref{zeroRPOfig:redudantPaths} shows the latency CDF (Cumulative Distribution Function) of the secondary paths in each network, considering only pairs of sites that replicate between each other. Results are shown for $\alpha = 0.05$, although other $\alpha$ values achieve close results and lead to the same conclusions. Note that all networks have several secondary paths with latency values much higher than the $L_{worst}$ parameter. In RNP, some paths have values close to 20~ms. For a better analysis, Figure~\ref{zeroRPOfig:percentComplyPaths} shows the fraction of secondary paths that meet the latency requirement $L_{worst}$. The X-axis groups the results for each analyzed network, for different $L_{worst}$ values. Note that the most stringent requirement (i.e., $L_{worst}=1.3$~ms), is met by 40\% of the paths in RENATER. Nevertheless, none of the paths in RNP and GEANT meet $L_{worst}=1.3$~ms. This better performance of RENATER is explained by the smaller region that this network spans, as compared with RNP and GEANT. For the most relaxed requirement, i.e., 5.6~ms, RENATER meets $L_{worst}$ in approximately 90\% of its paths. Despite these good results, the other networks, and even RENATER with more stringent requirements, have secondary paths with high latency values. This shows that the server placement alone is not enough to guarantee low latency values in secondary paths. Hence, the WAN topology should be modified to offer such service, especially in networks that span large geographical regions. This modification is related to the area of WAN design, which consists of choosing the network topology, link capacities and paths between nodes.
The literature on this type of problem is vast and generally addresses the design of optical networks~\cite{habib2013disaster}.
Our work focuses on server placement, considering that the WAN is already designed and installed. The joint optimization of the server placement and WAN design, suggested in~\cite{couto2014Network}, is a subject of future work.

As the possible secondary paths may have high latency values, the DC designer may choose to not configure these paths, setting $\gamma = 0$ when executing the problem formulated in Section~\ref{zeroRPOsec:problem}. For example, although important to guarantee service continuity, the proposal of SecondSite itself does not require the existence of these paths. Consequently, if secondary paths are not configured, more bandwidth is available to the installation of primary servers and their corresponding VM replications. On the other hand, if replication links are broken, the VMs on the primary servers should be paused to avoid the execution of operations without replication. However, this strategy reduces the availability of primary servers (i.e., the fraction of time that they are operational). Another strategy, employed by SecondSite, is to keep the primary servers running in the unprotected mode (i.e., operational but without replicating their operations)~\cite{rajagopalan2012secondsite}. This strategy reduces the RPO for the sake of availability.

\begin{figure}
\centering
\includegraphics[width=0.40\textwidth]{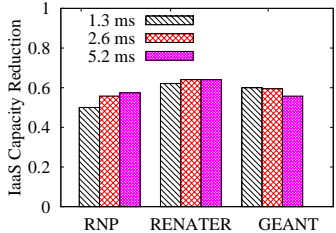}
\caption{Overhead of the secondary paths.}
\label{zeroRPOfig:comparisonWithoutRedundantServiceReduction}
\end{figure}

To analyze to which extent the configuration of secondary paths reduces the number of primary servers supported, we execute the optimization problem with the same parameter values used in the previous section, but setting $\gamma=0$. We denote as $S$ and $S'$, respectively, the number of primary servers supported in a network with secondary paths configured ($\gamma=1$) and when they are not used ($\gamma=0$). We thus evaluate the capacity reduction caused by secondary paths using the expression $1 -\frac{S}{S'}$. Figure~\ref{zeroRPOfig:comparisonWithoutRedundantServiceReduction} shows, for the different networks and latency requirements, the reduction when the fraction of allowed bandwidth ($\alpha$) is 0.05. Our results show that secondary paths significantly impact the service capacity, reducing at least 50\% of the primary servers for all results in Figure~\ref{zeroRPOfig:comparisonWithoutRedundantServiceReduction}. The results for other $\alpha$ values, omitted for the sake of conciseness, show the same behavior, always higher than 50\%. In a nutshell, this analysis shows that the DC designer can significantly increase the service capacity when choosing not to use secondary paths in the considered topologies.
%

\section{Related Work}
\label{zeroRPO:secRelacionados}

The server placement problem in a DC is a topic still incipient in the literature. Most of the contributions consider a traditional DC distribution~\cite{habib2012design,xiaoJoint2014,develder2011survivable,develder2012resilient,xiao2013Data}, adapted to scenarios such as Content Delivery Networks (CDNs)~\cite{pierre2006Globule}. Hence, they assume that the geo-distribution is achieved by the anycast principle, in which any node that runs a given service can reply to requests of this service. Consequently, these works do not discriminate backup and primary servers, since all servers in the network are operational at the same time.
In addition, these works do not consider the synchronization between servers, disregarding RPO (Recovery Point Objective) requirements. Finally, as their distribution scheme considers that all servers are operational,
they are not able to save server resources, as we do in this work. We detail next three other works that consider a scenario similar to our work.

In our previous work~\cite{couto2014Latency}, we analyze the trade-offs between latency and resilience when designing a geo-distributed DC. We show in that work the possibility to design a resilient DC, with servers spread in a region, without a significant latency increase caused by the large geographical distances between sites. However, the analysis is generic, not considering specific requirements of a given IaaS scenario. Hence, in this article we draw the attention to an IaaS cloud with zero RPO and analyze the behavior of this service according to its latency, bandwidth, and backup server efficiency requirements.

Yao~\textit{et al.}~\cite{yao2014minimizing} propose an optimization problem to choose backup sites in a geo-distributed DC. As their backup is based on periodic replications, the problem also schedules the time at which the servers perform their backups.
Hence, different from the continuous replication scheme considered in our problem, Yao~\textit{et al.} define backup windows. These windows are predefined intervals, at which the backups are sent between sites. Consequently, their service does not consider applications with zero RPO. The objective of their problem is to minimize the number of time intervals used by the backup windows or, in other words, the network capacity consumed by backups. As the backup is not continuous and does not require acknowledgment, that work disregard latency requirements.

Bianco~\textit{et al.}~\cite{bianco2010optimal} propose a placement problem similar to our work, which consists in allocating primary and backup resources to host VMs in an existent WAN (Wide Area Network).
The optimization problem chooses the primary disk for the VM and its corresponding backup disk, in such a way that both disks do not share the same site.
Bianco~\textit{et al.} thus propose three placement problems. The first one minimizes, for all sites, the number of hops between the site hosting the primary disk and the other site hosting the backup.
This optimization is an attempt to minimize the latency between the sites, since the disk synchronization is a latency-sensitive process. Note that Bianco~\textit{et al.} do not consider the propagation delay between sites, as we consider in this work, minimizing only the hop count. This approach may not be adequate to DCs spanning large geographical regions, in which the hop count is not necessarily related to latency.
In our work, we minimize hop count by allowing only one-hop neighbors to replicate VMs, and by limiting the maximum latency between these neighbors.
According to Bianco~\textit{et al.}, after the failure in the primary site, the process of VM migration is a highly intensive CPU task.
Given that, their second placement scheme minimizes the number of backup servers in a site, to minimize the CPU required per site. However, the global CPU capacity required (i.e., considering all the sites) to recover the VMs remains the same, regardless of the DC load distribution.
Note that this approach is the opposite of the backup sharing scheme proposed in our work, since our strategy aims to group as many backups as possible in the same site. Finally, the third problem is a hybrid approach, considering the hop count as well as the load balancing of backups.

Given the state of the art, the contribution of this article regarding the DC server placement consists of considering the continuous backup replication, which entail stringent latency and bandwidth requirement, and saving backup server resources.
Finally, we focus on an IaaS services, different from the traditional CDNs. 

Another area related to server placement in DCs is the survivable virtual networking embedding (SVNE), introduced for the first time by Rahman~\textit{et al.}~\cite{rahman2010survivable}. The virtual network embedding (VNE) consists of choosing which physical nodes and links are going to be used by the virtual networks requested. The survivable mapping chooses, in addition to the physical resources for normal operation, the physical resources for backups that will be used by the virtual network if a node or physical link fails. 
Rahman~\textit{et al.} consider only link failures and consider a fast restoration scheme, where the backup resources are reserved \textit{a priori}. Another SVNE algorithm, proposed by Yu~\textit{et al.}~\cite{yu2011cost}, consider the sharing of backup resources, reducing the amount of resources required. Another area related to our work is the placement of controllers in Software Defined Networks (SDNs). In SDN, the forwarding elements must be always reachable by a network controller. Hence, different works propose controller placement schemes, where the main goal is to guarantee that the forwarding elements have a path to at least one controller, in case of failures~\cite{muller2014survivor}. Finally, another related area is the resilient VM placement~\cite{bodik2012surviving}, where the algorithms try to distribute VMs in a DC to overcome the effect of failures (e.g., to eliminate single points of failure).

\section{Conclusions and Future Work}
\label{zeroRPO:conc}

In this work, we proposed a scheme to place servers in a geo-distributed DC supporting an IaaS (Infrastructure as a Service) cloud with zero RPO (Recovery Point Objective). This type of cloud has the challenge of requiring high bandwidth capacity and low latency values between the primary server and its corresponding backup.
Hence, we formulate an optimization problem with the goal to place as many as primary servers as possible, increasing the IaaS capacity. In addition, this problem takes advantage of the virtualization, which allows the sharing of backup servers, significantly reducing the number of installed servers. Results for all considered networks show that we can achieve at least 40\% of backup server efficiency. In comparison, classical schemes, where no backups are shared, present zero efficiency.
We also show that the efficiency can be even higher if we relax the latency requirement. This work also analyzes the properties of geo-distributed DCs if we configure secondary paths between pairs of sites.
These paths are employed if the replication link between the two sites fails. Results show that secondary paths can have high latency values, not meeting the replication service requirement. Finally, we show that, although the secondary paths are configured following a shared protection scheme, they require a high reserved bandwidth. Our results show, for all considered networks, that it would be possible to install at least twice the number of primary servers if we do not configure secondary paths.

As a future work, we plan to propose a DC design algorithm that jointly designs the WAN and places the servers. This can be useful, for example, to offer secondary paths with lower latency values. Another promising direction is to optimize the placement of servers that detect failures. Hence, the objective of this optimization problem would be to reduce the VM recovery time and to detect failures more accurately.

\section*{Acknowledgments}
The authors would like to thank FAPERJ, CNPq, CAPES research agencies and the ANR Reflexion project (contract nb: ANR-14-CE28-0019).




\bibliographystyle{model3-num-names}
\bibliography{comNetSIzeroRPO_vf}







\end{document}